\documentclass{elsart3}

\usepackage{graphicx}

\begin{document}

\begin{frontmatter}
\title{Low-energy excitations in a one-dimensional orthogonal dimer model with 
the Dzyaloshinski-Moriya interaction}

\author[a]{Kenji Sasaoka},
\author[a]{Chikara Ishii},

\address[a]{Depertment of Physics, Tokyo University of Science, 1-3 Kagurazaka, 
Shinjiku-ku, Tokyo, Japan}

\begin{abstract}
Effects of the Dzyaloshinski-Moriya (DM) interaction on low-energy excitations 
in a one-dimensional orthogonal-dimer model are studied by using the perturbation 
expansions and the numerical diagonalization method. In the absence of the DM 
interaction, the triplet excitations show two flat spectra with three-fold degeneracy, 
which are labeled by magnetization $M=0,\pm{1}$. These spectra split 
into two branches with $M=0$ and with $M=\pm{1}$ by switching-on of the DM 
interaction and besides the curvature appears in the triplet excitations 
with $M=\pm 1$ more strongly than those of $M=0$. 
\end{abstract}

\end{frontmatter}

\section{Introduction}

Since its discovery by Kageyama {\it et al}.\cite{prl1}, the spin dimer 
compound SrCu$_2$(BO$_3$)$_2$ has attracted much attention as a suitable material 
for frustrated spin systems in low dimension. SrCu$_2$(BO$_3$)$_2$ exhibits various 
interesting phenomena, such as a quantum disordered ground state~\cite{prl1} and a complex shape 
of magnetization curve\cite{jpsj1}, because of its unique crystal structure.  
In consideration of the structure, Miyahara and Ueda suggested that the magnetic 
properties of the spin dimer compound SrCu$_2$(BO$_3$)$_2$ can be described as a 
spin-$1/2$ two-dimensional (2D) orthogonal-dimer model~\cite{prl2}, equivalent to  the 
Shastry-Sutherland model on square lattice with some diagonal bonds~\cite{PhysicaB1}. 
The ground state of the Shastry-Sutherland model in dimer phase is exactly 
represented by a direct product of singlets. 
The low-energy dispersions possess six-fold degeneracy
and are almost flat reflecting that the triplet tends to localize on vertical or 
horizontal bonds.  
\par

Recent experiments by ESR~\cite{Nojiri} and neutron inelastic scattering~(NIS) 
have observed splitting of degenerate dispersions of SrCu$_2$(BO$_3$)$_2$,
which can not be explained by the {\it isotropic} Shastry-Sutherland model.
Hence C{\' e}pas {\it et al.} pointed out that the Dzyaloshinski-Moriya~(DM) 
interaction~\cite{DM} must be added between vertical and horizontal dimers in the isotropic 
Shastry-Sutherland model in order to explain the splitting.~\cite{prl3}
\par

In this paper, as a simple model to clarify effects of the DM interaction to 
low-energy excitations in orthogonal-dimer systems, one-dimensional (1D) 
orthogonal-dimer model with the DM interaction is studied by using the perturbation 
theory and the numerical exact-diagonalization method. In the absence of the 
DM interactions, properties of ground state, low-energy excitations, and 
magnetization processes of the 1D orthogonal dimer model has been studied by 
several authors.~\cite{prb1,prb2,prl4}  
\par

\section{Model}
The Hamiltonian of the 1D orthogonal-dimer model with the DM interaction is 
given by
\begin{eqnarray}
\mathcal{H} = \mathcal{H}_{\mathrm{intra}} +\mathcal{H}_{\mathrm{inter}} 
+ \mathcal{H}_{\mathrm{DM}}, 
\label{eq:Hamiltonian}
\end{eqnarray}
where
\begin{eqnarray}
\mathcal{H}_{\mathrm{intra}}&=&J \sum _{j=1}^{N} 
   \left({\mbox{\boldmath$S$}}_{j,1}
   \cdot{\mbox{\boldmath$S$}}_{j,2}+{\mbox{\boldmath$S$}}_{j+\frac{1}{2},1}\cdot
   {\mbox{\boldmath$S$}}_{j+\frac{1}{2},2}\right), 
\label{eq:intra}\\
\mathcal{H}_{\mathrm{inter}}&=&J'\sum _{j=1}^{N} 
   \left\{ \left({\mbox{\boldmath $S$}}_{j,1}+{\mbox{\boldmath$S$}}_{j,2}\right)
   \cdot{\mbox{\boldmath $S$}}_{j+\frac{1}{2},1}\right.
   \nonumber \\
&&+\left.{\mbox{\boldmath$S$}}_{j+\frac{1}{2},2}\cdot
   \left({\mbox{\boldmath$S$}}_{j+1,1}+{\mbox{\boldmath$S$}}_{j+1,2}\right) 
   \right\}, 
   \label{eq:inter}\\
\mathcal{H}_{\mathrm{DM}}&=&
   {\mbox{\boldmath$D$}}\cdot\sum_{j=1}^{N}\left(
   {\mbox{\boldmath$S$}}_{j+\frac{1}{2},1}\times{\mbox{\boldmath $S$}}_{j,1}+  
   {\mbox{\boldmath$S$}}_{j,1} \times{\mbox{\boldmath $S$}}_{j-\frac{1}{2},2}\right. 
   \nonumber \\
&&+\left.{\mbox{\boldmath $S$}}_{j-\frac{1}{2},2}\times{\mbox{\boldmath$S$}}_{j,2}+
   {\mbox{\boldmath$S$}}_{j,2} \times{\mbox{\boldmath$S$}}_{j+\frac{1}{2},1}\right). 
   \label{eq:DM}
\end{eqnarray}
Here $N$ is the number of unit cells in the system, as shown by a broken rectangle 
in Fig.~1. The unit cell includes two dimers along vertical and horizontal direction,
which are designated by the index, $j$ and $j+\frac{1}{2}$, respectively. 
${\mbox{\boldmath $S$}}_{j,\alpha}$ ($j=1,2,\cdots,N$ and $\alpha=1,2$) denotes a 
spin-$1/2$ operator on $\alpha$-spin in $j$-th dimer. $J$ and $J'$ severally indicate 
the exchange coupling in intra-dimer and in inter-dimer. Due to the structure of system,
the DM exchange interaction, ${\mbox{\boldmath $D$}}$, exists only on inter-dimer bonds 
and has only a component perpendicular to two kinds of dimer in the unit cell. The 
periodic boundary condition is imposed to the system.
 
\begin{figure}[t]
  \begin{center}
    \includegraphics[keepaspectratio=true,width=75mm]{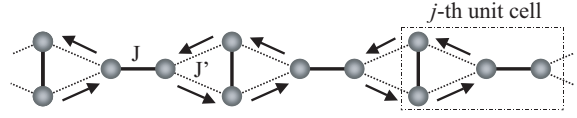}
  \end{center}
  \caption{The one-dimensional orthogonal-dimer model with the DM interaction. The thick 
  and broken lines indicate intra- and inter-dimer interactions, respectively. The arrows 
  on inter-dimer show the order of spins in the expression of the DM interaction
  ${\mbox{\boldmath $D$}}\cdot({\mbox{\boldmath $S$}}_{i,\alpha}\times{\mbox{\boldmath $S$}}_{j,\beta})$,
  that is $(i,\alpha)\to(j,\beta)$. The unit cell includes a vertical and horizontal dimer.
  The former dimers are at $j$-site and the latter at $\left(j+\frac{1}{2}\right)$-site.}
  \label{f1}
\end{figure}

\section{Ground state and low-energy excitations}
In this section, let us discuss the ground state and low-energy excitations of the 1D 
orthogonal dimer model with the DM interaction. We can expect that the ground state is 
in the dimer phase in the limit of strong intra-dimer coupling ($J\gg J',|{\mbox{\boldmath $D$}}|$), 
even when the DM interaction is switched on the isotropic system. Therefore, it is reasonable to 
treat the intra-dimer Hamiltonian~(\ref{eq:intra}) as an unperturbated one and the others as 
perturbation. 

The inter-dimer interaction $J'$ creates two adjacent triplets from a pair of a singlet and 
triplet and vice versa, and besides shows scatterings between two triplets. The DM interaction 
not only causes the former process but also creates or annihilates two adjacent singlets.   
Therefore the DM interaction can play a crucial role in the ground state and the low-energy 
excitations in the dimer phase.
\par

First, 
we discuss the ground-state energy of Hamiltonian~(\ref{eq:Hamiltonian}).
In the absence of the DM interaction, the ground state is exactly represented by a direct 
product of singlets and its energy is given as $\epsilon_0\equiv{E_0/J}=-\frac{3}{2}N$. 

On the other hands, the ground-state energy of total Hamiltonian~(\ref{eq:Hamiltonian})
is estimated as
\begin{eqnarray}
\frac{E_g}{J}=\epsilon_0-\frac{N}{2}\left(\frac{D}{J}\right)^2
-\frac{N}{8}\frac{J'}{J}\left(\frac{D}{J}\right)^2.
\label{eq:groundE}
\end{eqnarray}
from the perturbation expansion up to the third order in $J'/J$ and $D/J$.
The result means that the ground state cannot be exactly described by the direct product 
of singlets owing to the DM interaction.
\par

Next, we argue the low-energy excitations in the system. Since the ground state belongs 
to the dimer phase in the region of strong-$J$, the lowest excited states will be well 
described by
\begin{eqnarray}
&&|M, k \rangle _{\rm ver}\equiv  t_{M,k}^{{\rm ver} \, \dagger} |0 \rangle
\nonumber\\
&&=\frac{1}{\sqrt{N}}\sum_{j=1}^{N} \mathit{e}^{ikj} | s_1 
\cdots t_{M,j} \cdots s_{N+\frac{1}{2}}\rangle, \label{eq:vFourier}\\
&&|M, k \rangle _{\rm hor} \equiv t_{M,k}^{{\rm hor} \, \dagger} |0 \rangle \nonumber\\
&&=\frac{1}{\sqrt{N}}\sum_{j=1}^{N} 
\mathit{e}^{ik\left(j+\frac{1}{2} \right)}
|s_1\cdots t_{M,j+\frac{1}{2}}\cdots s_{N+\frac{1}{2}}\rangle.
\label{eq:pFourier}
\end{eqnarray}
Here, $M(=0,\pm{1})$ and $k$ are the total magnetization and the wave number, respectivery.
$s_j$ and $t_{M,j}$ in ket severally denote a singlet and a triplet with $M$ at $j$-site and, 
$t_{M,k}^{{\rm ver} \, \dagger}$~($t_{M,k}^{{\rm hor} \, \dagger}$) is defined as an operator to create a triplet 
propagating on vertical~(horizontal) dimers.
By using two states of Eqs.~(\ref{eq:vFourier}) and (\ref{eq:pFourier}), the Hamiltonian (1) is projected on following ($2\times{2}$)-matrix:
\begin{eqnarray}
\mathcal{H}_{M}(k)={\mbox{\boldmath $v$}}_M^\dagger(k){\mbox{\boldmath $A$}}_M(k){\mbox{\boldmath $v$}}_M(k) 
\label{eq:hm}
\end{eqnarray}
where
\begin{eqnarray}
{\mbox{\boldmath $A$}}_M(k) \equiv 
\left[
  \begin{array}{cc}
    a_{11,M}   & a_{12,M}   \\
    a_{21,M}   & a_{22,M}   \\
  \end{array}
\right],~
{\mbox{\boldmath $v$}}_M(k)\equiv
\left[
  \begin{array}{c}
    t_{M,k}^{\rm ver}   \\
    t_{M,k}^{\rm hor}  \\
  \end{array}
\right].
\end{eqnarray}

The Eq.~(\ref{eq:hm}) for $M=0$ has no off-diagonal elements within perturbation up to the third order. 
Therefore the excitation energies for $M=0$ are given by 
\begin{eqnarray}
\frac{E_{0}^{{\rm ver}}\left(k\right)}{J}&=&\epsilon_0+1
-\frac{N-1}{2}\left(\frac{D}{J}\right)^2-\left(\frac{J'}{J}\right)^2\nonumber\\
& &-\frac{N-1}{8}\frac{J'}{J}\left(\frac{D}{J}\right)^2
-\frac{1}{2}\left(\frac{J'}{J}\right)^3,
\label{eq:excitedE0}\\
\frac{E_{0}^{{\rm hor}} \left( k \right)}{J} &=& \epsilon_0 +1 -\frac{N+1}{2}\left(\frac{D}{J}\right)^2 \nonumber\\
& &-\left(\frac{N+3}{8}+\cos{k} \right)\frac{J'}{J}\left(\frac{D}{J}\right)^2.
 \label{eq:excitedE1}
\end{eqnarray}
In contrast to the 2D orthogonal dimer model, two excitation energies, $E_{0}^{{\rm ver}}\left(k\right)$ and $E_{0}^{{\rm hor}} \left( k \right)$, split in the case of 1D system. It is also interesting to note that the curvature of $E_{0}^{{\rm hor}} \left( k \right)$ appears in the third ordered correction in Eq.~(\ref{eq:excitedE1}).
\par

On the other hand, the projected Hamiltonian with $M=\pm1$ 
has an off-diagonal element. The perturbation calculation up to the third order leads to the matrix:
\begin{eqnarray}
{\mbox{\boldmath $A$}}_{\pm1}=
\left[
  \begin{array}{cc}
    a+b_k   & -i d_k   \\
    id_k    & a+c   \\
  \end{array}
\right],
\label{eq:apm1}
\end{eqnarray}
where
\begin{eqnarray*}
a  &=& \epsilon_0 +1 -\frac{N-1}{2}\left(\frac{D}{J}\right)^2-\left(\frac{N}{8}-\frac{3}{16} \right)\frac{J'}{J}\left(\frac{D}{J}\right)^2,\\
b_k&=&-\left(\frac{J'}{J}\right)^2-\frac{1}{4}\left(\frac{D}{J}\right)^2 \cos k -\frac{1}{2}\left(\frac{J'}{J}\right)^3\\
   & &-\frac{1}{2}\left(1+\cos k \right)\frac{J'}{J}\left(\frac{D}{J}\right)^2,\\
c  &=&-\frac{1}{2}\left(\frac{D}{J}\right)^2,\\
d_k&=&
\left(\frac{D}{J}+\frac{J'D}{2J^2}-\frac{J'^{2}D}{4J^3}-\frac{J'D^2}{4J^3}  
  + \frac{D^3}{2J^3}\right)\cos\frac{k}{2}. 
\end{eqnarray*}
By diagonalizing Eq.~(\ref{eq:apm1}), 
the excitation energies with $M=\pm1$ are 
obtained as
\begin{eqnarray}
\frac{E_{\pm1}^{\left(\pm\right)}\left(k\right)}{J} = a+\frac{b_k+c}{2} \pm \sqrt{d_k^2+\frac{\left( b_k-c \right)^2}{4}}.
 \label{eq:excitedE3}
\end{eqnarray}
The curvature of $E_{\pm1}^{\left(\pm\right)}\left(k\right)$ is dominant by the first ordered correction with regard to $D/J$ in the off-diagonal element $d_k$. The correction derives from the scattering between a singlet and a triplet with $M=\pm{1}$ due to the DM interaction.
\par

Subtracting the ground-state energy of Eq.~(\ref{eq:groundE}) from excited-state energies of Eq.~(\ref{eq:excitedE0}), (\ref{eq:excitedE1}), and (\ref{eq:excitedE3}), 
the low-energy dispersions, $\omega_{k,M}\equiv(E_M(k)-E_g(k))/J$, are estimated as  
\begin{eqnarray}
\omega_{0}^{\rm ver}\left(k\right)&=& 1 + \frac{1}{2}\left(\frac{D}{J}\right)^2 -\left(\frac{J'}{J}\right)^2\nonumber\\ 
& &+ \frac{1}{8}\frac{J'}{J}\left(\frac{D}{J}\right)^2 - \frac{1}{2}\left(\frac{J'}{J}\right)^3,
 \label{eq:omega0} \\
\omega_{0}^{\rm hor}\left(k\right)&=& 1-\frac{1}{2}\left(\frac{D}{J}\right)^2
-\left(\frac{3}{8}+\cos{k}\right)\frac{J'}{J}\left(\frac{D}{J}\right)^2,
\nonumber\\
\label{eq:omega1}\\
\omega_{\pm1}^{(\pm)}\left(k\right) &=& 1 + \frac{1}{2}\left(\frac{D}{J}\right)^2 -\left(\frac{J'}{J}\right)^2+ \frac{3}{16}\frac{J'}{J}\left(\frac{D}{J}\right)^2 
\nonumber\\
& & +\frac{b_k+c}{2} \pm \sqrt{d_k^2+\frac{\left( b_k-c \right)^2}{4}}.
\label{eq:omega2}
\end{eqnarray}

Figure 2 shows the low-energy dispersions for $J'/J=0.4$ and $D/J=0.2$. The low energy 
spectra $\omega_{0}^{\rm ver}$ and $\omega_{0}^{\rm hor}$ are severally 
represented by the lower and upper solid lines, 
and then the upper and lower dotted lines denote $\omega_{\pm1}^{(+)}$ and  
$\omega_{\pm1}^{(-)}$ in Eqs.~(\ref{eq:omega2}).
The full and open circles represent the low-energy spectra with $M=0$ and $M=\pm{1}$.
The perturbation theory is in agreement with the numerical diagonalization, as to
low-energy excitations.
The dispersions with same $M$ are not degenerate, which 
happens even if the DM interaction is not taken account of. 
Therefore the inter-dime coupling $J'$ is also important for splitting of the low-energy 
dispersions with same $M$.
This is because the parity on the vertical dimer is conserved in the one-dimension system 
without the DM interaction.
The DM interaction not only splits into branches with $M=0$ and $M=\pm{1}$,
but also makes triplets move more easily.
\begin{figure}[tb]
  \begin{center}
    \includegraphics[keepaspectratio=true,width=70mm]{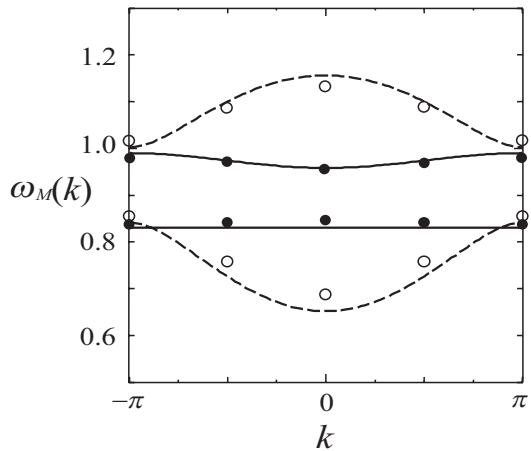}
  \end{center}
\caption{
The low-energy dispersions for $J'/J=0.4$ and $D/J=0.2$.
The solid and dotted curves are of $M=0$ and $M=\pm1$, respectively.
The full~(open) circles indicates the excitation energies for 
$M=0$~($M=\pm{1}$) calculated by numerical diagonalizations.}
\label{f2}
\end{figure}
\par

\section{Summary}
We investigated the low-energy excitations 
in the 1D orthogonal-dimer model with the DM interaction using the perturbation theory and 
numerical exact-diagonalization method. 
The DM interaction allows a triplet to propagate in singlet sea as seen in the 2D system~\cite{jpsj2}, while the triplet is localized on vertical or horizontal dimer in the absence of the DM interaction. 
This curvature effect happens in 
a two-dimensional orthogonal-dimer model, but the splitting of spectra reflects a stringent 
constraint for the motion of a triplet due to one dimensionality as well as the DM interaction.

\begin{ack}
We gratefully acknowledge helpful discussions and comments with Tetsuro Nikuni, Masaaki Nakamura, and Takahiro Yamamoto on several points in this work.
We would also like to thank K$\rm{\hat o}$ichir$\rm{\hat o}$ Ide for valuable advice on numerical techniques.
\end{ack}

\end{document}